\documentclass[article,pre,
twocolumn,
amsmath,amssymb,
amsmath,amssymb, aps,
]{revtex4}
\usepackage[dvipsnames]{xcolor} 
\usepackage{epsfig}
\usepackage{amsfonts}
\usepackage{amsmath}
\usepackage{wasysym}
\usepackage{amssymb}
\usepackage{bm} 
\usepackage[framemethod=PStricks]{mdframed}
\usepackage{lipsum}
\usepackage{float}
\usepackage{enumitem}

\usepackage{pdflscape}
\usepackage{graphicx}
\usepackage{ulem}

\newcommand{\be}{\begin{equation}}
\newcommand{\ee}{\end{equation}} 
\newcommand{\bea}{\begin{eqnarray}} 
\newcommand{\eea}{\end{eqnarray}}
\newcommand{\bfr}{{\bf{r}}}
\newcommand{\bfk}{{\bf{k}}}
\newcommand{\bfv}{{\bf v}}
\newcommand{\bfnabla}{{\boldsymbol \nabla}}
\newcommand{\lr}[1]{\left(#1\right)}
\usepackage[utf8]{inputenc}
\usepackage{bbm} 				% identity operator
\usepackage{epstopdf}				% self explanatory
\usepackage{verbatim}    			% long comments
\usepackage{sidecap}                % captions on the side
\usepackage{indentfirst}
\usepackage[colorinlistoftodos]{todonotes}

\definecolor{armygreen}{rgb}{0.0, 0.5, 0.0}

\begin{document}
%\linenumbers

\title{Circulation Statistics and the Mutually Excluding \\ Behavior of Turbulent Vortex Structures}

\author{L. Moriconi$^{1}$, R.M. Pereira$^{2}$\footnote{corresponding author: rpereira@mail.if.uff.br}, and V.J. Valadão$^{1}$}
\affiliation{$^{1}$Instituto de F\'\i sica, Universidade Federal do Rio de Janeiro, \\
C.P. 68528, CEP: 21945-970, Rio de Janeiro, RJ, Brazil}
\affiliation{$^{2}$Instituto de Física, Universidade Federal Fluminense, 24210-346 Niterói, RJ, Brazil}

%\date{}

\begin{abstract}
The small-scale statistical properties of velocity circulation in classical homogeneous and isotropic turbulent flows are assessed through a modeling framework that brings together the multiplicative cascade and the structural descriptions of turbulence. We find that vortex structures exhibit short-distance repulsive correlations, which is evidenced when they are ``tomographically" investigated, by means of planar cuts of the flow, as two-dimensional vortex gases. This phenomenon is suggested from model improvements which allow us to obtain an accurate multiscale description of the intermittent fluctuations of circulation. Its crucial new ingredient, the conjectured hard disk behavior of the effective planar vortices, is then found to be strongly supported from a study of their spatial distributions in direct numerical simulations of the Navier-Stokes equations.
\end{abstract}

%\pacs{}

\maketitle

Velocity circulation, the subject of some of the most celebrated theorems of fluid dynamics \cite{chorin1}, is a central concept that pervades a broad spectrum of phenomena in classical and quantum fluids \cite{lugt,barenghi_etal}. Circulation phenomenology is expected to be particularly relevant in turbulent systems. Actually, it has been long suggested, from the visualization of intense vorticity domains in direct numerical simulations (DNS) \cite{orszag_etal,farge_etal,kaneda_etal}, that homogeneous and isotropic turbulence could be effectively depicted as a dilute gas of long-lived vortices.

It was not until a few years ago, however, that hardware improvements in high performance computing platforms have finally allowed the implementation of extensive numerical simulations, necessary for a deeper scrutiny of turbulent circulation \cite{Iyer_etal1,Iyer_etal2}. As a result, novel theoretical and phenomenological accounts of circulation intermittency have been subsequently developed \cite{migdal,apol_etal,mori,mori_PNAS,mori_pereira}, including closer connections between classical and quantum turbulence \cite{muller_etal1, polanco_etal,muller_etal2}.

We focus on a recent modeling framework of circulation statistics which unifies both the structural (turbulence seen as a vortex gas) and multiscale (turbulence seen as a multiplicative cascade process) aspects of turbulent flows \cite{apol_etal,mori,mori_pereira}. Our aim is to predict statistical features of turbulent circulation from the superimposed contributions of individual thin vortex tubes. This task will be carried out here along the lines of Monte Carlo simulations, which will allow us not only to subject the vortex gas model
to a rigorous validation test, but also to find unsuspected phenomenological results, which otherwise would be of very (if not prohibitive) difficult analytical reach.

To start, let $\mathcal{D}$ be an oriented bounded region of area $A$ contained in a plane $\gamma$ that ``slices" the entire turbulent domain. The velocity circulation around its contour can be expressed, in the vortex gas model, as
\be
\Gamma  (\mathcal{D}) = \sum_i \Gamma_i (\mathcal{D}) \ , \  \label{G_C}
\ee
where the $\Gamma_i$ are the circulation contributions conveyed by each of one of the vortex tubes that cross $\gamma$. All one needs to describe fluctuations of $\Gamma (\mathcal{D})$, therefore, is a statistical model for the joint random variables $\Gamma_i$.
To this end, we write, following the guidelines of \cite{apol_etal,mori}, and in appropriate units of circulation,
\be
\Gamma_i (\mathcal{D}) = \xi_{\mathcal{D}} \int_\mathcal{D} d^2 \bfr  g_\eta(\bfr-\bfr_i) \tilde \omega (\bfr_i)  \ , \ \label{gamma_i}
\ee
where
\begin{enumerate}[wide, labelindent=0pt,topsep=3pt,itemsep=1pt,parsep=2pt,label=(\roman*)]

\item \mbox{$\bfr_i \in \gamma$} stands for the center position of a two-dimensional vortex structure (the intersection of a vortex tube with the cutting plane $\gamma$);

\item the Gaussian envelope \mbox{$g_\eta(\bfr) \equiv \exp [ - \bfr^2 / (2 \eta^2) ]$} models the vorticity decay of planar vortices, assumed to have core radius \mbox{$\eta \equiv a \eta_K$}, where $a$ is a dimensionless constant (an input modeling parameter) and $\eta_K$ is the Kolmogorov dissipative length scale \cite{frisch};

%\item $\tilde \omega (\bfr)$ is a bounded Gaussian random field with two-point correlation function $\langle \tilde \omega (\bfr) \tilde \omega (\bfr') \rangle  \sim 1/|\bfr - \bfr'|^\alpha$ in the inertial range, and \mbox{$\alpha = 4/3 - \mu/4$}, where $\mu \simeq 0.17$ is the intermittency exponent derived from the scaling behavior of the energy dissipation rate field $\epsilon (\bfr)$ \cite{tang_etal};

\item $\tilde \omega (\bfr)$ is a Gaussian random field whose two-point correlation function behaves as
$\langle \tilde \omega (\bfr) \tilde \omega (\bfr') \rangle  \allowbreak \sim 1/|\bfr - \bfr'|^\alpha$ in the inertial range and \mbox{$\alpha = 4/3 - \mu/4$} \cite{mori}, where $\mu \simeq 0.17$ is the intermittency exponent derived from the scaling behavior of the energy dissipation rate field $\epsilon (\bfr)$ \cite{tang_etal}. We note that $\tilde \omega (\bfr)$ must be regularized over the scale $\eta_K$, such that its variance remains finite.

\item $\xi_\mathcal{D} \equiv (1/A)\int_\mathcal{D} d^2 \bfr \sqrt{\epsilon(\bfr)/\epsilon_0}$, with $\epsilon_0$ being the mean dissipation rate (conveniently set to unity),
is a modulating random field for the amplitude of circulation fluctuations over $\mathcal{D}$.
Its square root dependence upon the dissipation field is closely related to the similarity
hypotheses that have been previously put forward to model velocity gradient fluctuations \cite{wyn_tenne,kholmy_etal}.
\end{enumerate}

%In consonance
According to the Gaussian multiplicative chaos description of the
turbulent cascade \cite{ro_va}, a field-theoretical extension of the Obukhov-Kolmogorov (OK62)
theory of intermittency \cite{O62,K62}, it is implied that the two-dimensional measure $\xi_D$ behaves as a lognormal variable as well, with
%we take $\xi_\mathcal{D}$ to be a lognormal
%random variable, with
\be
\ln(\xi_\mathcal{D}) \sim {\cal{N}}(-X_\mathcal{D},X_\mathcal{D}) \label{XD} \ , \
\ee
where, for a flow with Taylor based Reynolds number $R_\lambda$,
\be
X_\mathcal{D} = \frac{3 \mu}{8} \ln \left [ \frac{R_\lambda}{\sqrt{15}} \left ( \frac{\eta_K}{b r + \eta_K} \right )^{\frac{2}{3}} \right ] \ . \  \label{Xxi}
\ee
Above, $r$ yields a length scale for the domain $\mathcal{D}$ and $b$ is another phenomenological dimensionless parameter. Throughout this Letter, we take $\mathcal{D}$ to be a square domain of side $r$.

To completely set the vortex gas model, besides the parameters $a$ and $b$ just introduced, we have to prescribe the way the vortex structures are randomly distributed over $\gamma$. In Bayesian language, we take a prior distribution of planar vortices derived from a Poissonian point process characterized by the surface density field $\sigma (\bfr) = \sum_{i=1}^N \delta^2(\bfr -\bfr_i)$ with $\langle \sigma (\bfr) \rangle = \bar \sigma$. In this way, \mbox{Eq.~(\ref{G_C})} can be rewritten, by means of Eq.~(\ref{gamma_i}), as
\be
\Gamma (\mathcal{D}) =\xi_\mathcal{D} \int_{\mathcal{D}} d^2 \textbf{r}\int_\gamma d^2 \textbf{r}' g_\eta(\textbf{r}-\textbf{r}')\tilde{\omega}(\textbf{r}')\sigma(\textbf{r}') \label{gamma_3} \ . \
\ee

It is fundamentally important, concerning practical matters, to devise a systematic procedure for the determination of the parameters $a$, $b$ and $\bar\sigma$. In the original model \cite{apol_etal}, this is accomplished by using
the dilute gas approximation to analytically compute the circulation
flatness over a circular contour of radius $R$, $\mathcal{F}_R \equiv \langle\Gamma_R^4\rangle/\langle\Gamma_R^2\rangle^2$, in the $R \ll \eta_K$ and $R \gg \eta_K$ limits. This computation is a particular case of the higher order development put forward in Sec.~I of the Supplemental Material \cite{SM} when only the lowest order terms are retained. Comparisons can then be done to results derived from DNS data. Working in the small $R$ limit, we get the Reynolds number dependence of $\bar\sigma$ by matching the computed $\mathcal{F}_R$ with the empirical power law
\be
\lim_{R\to 0}\mathcal{F}_R \sim C_4 R_\lambda^{\alpha_4}  \ , \ \label{FR}
\ee
with $C_4 \simeq 1.16$ and $\alpha_4 \simeq 0.41$,
observed from the DNS data of Ref.~\cite{Iyer_etal1}. It turns out that
\be
\bar\sigma\pi\eta^2 = \frac32 \frac{1}{C_4}\frac{1}{15^\frac{3\mu}{4}}R_\lambda^{\frac{3\mu}{2}-\alpha_4}
\ , \ \label{barsigma}
\ee
which actually defines the expected number of vortices in a disk of radius $\eta$.

The parameter $a$ controls the initial curvature of $\mathcal{F}_R$ as $R$ increases, while its large $R$ limit is fixed by $b$. We refer the reader to \cite{apol_etal} for details, where the Reynolds number independent parameters $a=3.3$ and $b=2.0$ were established as a first approximation.

Even though it is possible to work out analytical expressions for the circulation flatness in the limits of small and large domains, the same does not hold for intermediate ranges. To cope with this issue, we have performed Monte Carlo simulations of Eq.~(\ref{gamma_3}) to numerically evaluate the circulation flatness at various length scales.

In order to produce Monte Carlo statistical ensembles out of Eq.~(\ref{gamma_3}), where $\sigma(\bfr)$ is straightforwardly drawn from a Poissonian point process, samples of $\xi_{\mathcal{D}}$ are easily generated from the prescriptions (\ref{XD}) and
(\ref{Xxi}). The field $\tilde\omega (\bfr)$ is, on its turn, realized as a two-dimensional
%fractional Brownian field \cite{heneghan_etal}
long-range correlated Gaussian random field regularized at the scale $\eta_K$. %This means, for the sake of clarity, that
Numerically, this is accomplished as \cite{saupe,stanley92,javerzatetal20}
% \be
% \tilde \omega ( \bfr ) = \int d^2 \bfk \, \psi (\bfk) k^{\frac{\alpha}{2}-1} \exp \left ( i \bfk \cdot \bfr - k \frac{\eta_K}{2} \right ) \ , \ \label{fbm}
% \ee
% where $\psi (\bfk)$ is a delta-correlated Gaussian random field in Fourier space.
\be
\tilde \omega ( \bfr ) = \frac{1}{C_{\eta_K}} \sum_{\bfk} \, \hat\psi (\bfk) k^{\frac{\alpha}{2}-1} \exp \left ( i \bfk \cdot \bfr - k \frac{\eta_K}{2} \right ) \ , \ \label{fbm}
\ee
where the $\hat\psi(\bfk)$ are random uncorrelated Gaussian variables, $C_{\eta_K}$ is a normalization constant to ensure unit variance and the sum is taken over the three components of $\bfk$ such that $k=|\bfk|\neq 0$.
Our numerical computations have been performed over grids having the same resolution parameters as the DNS data to which they are compared. As for those, we use both data from Ref.~\cite{Iyer_etal1} ($R_\lambda = 240$, 650, and 1300) as well as data processed from the Johns Hopkins Turbulence Databases \cite{JHTD,JHTD2} ($R_\lambda = 433$ and 610).

\begin{figure}[t]
\includegraphics[width=0.49\textwidth]{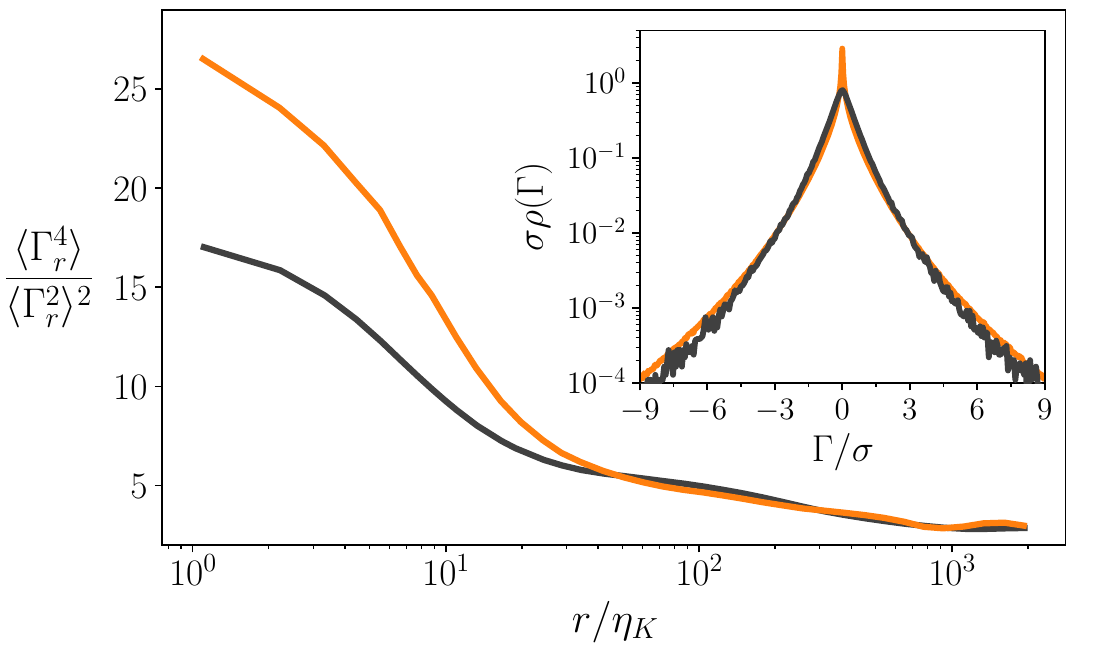}
\caption{Comparison between Monte-Carlo simulations of the vortex gas model (orange/gray) and DNS data (black) for $R_\lambda = 610$. Main frame: circulation flatness evaluated on square contours of side $r$. Inset: standardized cPDF for $r=2.2\eta_K$.}
\label{fig:poisson}
\end{figure}

Unexpectedly, Monte Carlo simulations reveal a mismatch with DNS results. We report, in Fig.~\ref{fig:poisson}, the scale dependence of the circulation flatness for $R_\lambda=610$. While the model closely follows the DNS curve for large $r$ (down to $r\sim 40\eta_K$), its circulation fluctuations become too intermittent for smaller contours, with a much larger flatness than expected. This is translated into sharply peaked circulation probability distribution functions (cPDFs), in contrast to those obtained from DNS, as shown in the inset for $r=2.2\eta_K$. A refreshed critical analysis of the vortex gas model is therefore in order.

One could suspect that higher order contributions to the dilute gas approximation could lead to non-negligible corrections to the circulation flatness associated to small contours. In other words, that the former expressions (\ref{FR}) and (\ref{barsigma}), used to fix $a$ and $\bar\sigma$, should be improved from the evaluation of further terms in the perturbative expansions of circulation moments. However, as it is shown in the Sec.~I of the Supplemental Material \cite{SM}, the Poissonian model of localized vortex structures gives subdominant corrections which would drive us to values of $\bar\sigma$ that are not consistent with the assumption that vortex tubes form a dilute system, once their cores are not observed to overlap at all. In any case, we empirically inspected different values for the parameters and found that no combination was able to address the issues observed in Fig.~\ref{fig:poisson}.

In short, the Poissonian vortex gas model is unable to describe the small scale distribution of vortex structures. Its paradoxical success at the lowest level of perturbation theory suggests, nevertheless, that the related subdominant contributions could be suppressed (or attenuated) from the consideration of additional/alternative modeling physical ingredients. A relatively simple variation of the Poissonian spatial distribution that incorporates this idea is the one provided by a gas of hard disks, which effectively introduces a small scale repulsive interaction between vortices by preventing their centers to come any closer than twice their radius.
As a matter of fact, excluded volume effects between vortices are not completely extraneous in turbulence modeling. They play an important role in refinements of the attached eddy description of turbulent boundary layers
\cite{silva_etal}.

In a statistical sense, the existence of short-distance vortex repulsion should not be very surprising, since small scale clusters of thin vortex tubes are likely to be polarized \cite{burger_etal} and, thus, subject to energy barriers against densification. It is interesting to note that small scale vortex polarization has been numerically observed (and quantified) in the context of quantum turbulence \cite{polanco_etal}. As it will be made clear next, our expectations are fulfilled by a hard disk model of planar vortices, which not only leads to excellent results for the circulation statistics, but also
reveals intriguing features on the statistical behavior of vortex structures.

The generation of hard disk ensembles of maximized entropy is a fundamental and challenging topic in statistical mechanics \cite{metropolis,alder62,isobe16} and the development of efficient algorithms has been an exciting field of research in
recent years \cite{isobe99,krauth1,andersonetal13,krauth2,krauth3}. We benefit, for our statistical analyses, on the publicly available code reported in Refs.~\cite{krauth4,krauth3}.

\begin{figure}[ht]
\includegraphics[width=0.49\textwidth]{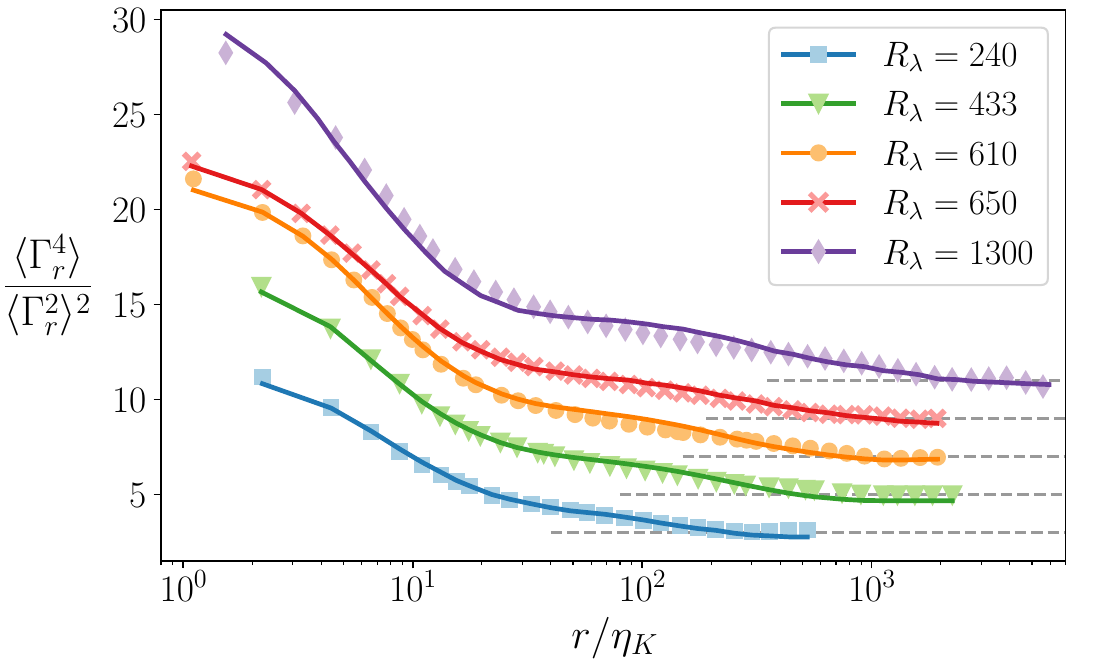}
\caption{The circulation flatness evaluated on square contours of side $r$ at different Reynolds numbers. Curves are vertically displaced (by steps of two units) for clarity. Symbols: vortex gas model with hard disks. Lines:
DNS. Dashed lines: the (also displaced) reference values for a Gaussian process.}
\label{fig:harddisksflat}
\end{figure}

Hard disk gases can be studied to some extent as perturbed Poisson distributions \cite{saundersetal,baddeleynaira,baddeleynairb}, but evaluations related to higher order statistics are usually tricky. Taking into account that perturbative deviations should not be too large, as the vortex gas is not dense, we adopt the more pragmatic point of view of inspecting values of $a$, $b$, and $\bar \sigma$ around the (dominant) Poissonian ones, looking for the best fitting results for the curves of circulation flatness.

Independent optimization fits give $a=3.3$ and $b$ in the range $1.6-1.8$ for all of the studied Reynolds number cases.
The optimal values of $\bar\sigma \pi \eta^2$ found for $R_\lambda$ = 240, 433, 610, 650, and 1300 are approximately $0.35$, $0.31$, $0.29$, $0.31$, and $0.29$ which are again close to those predicted by the dilute gas approximation, ($0.39$, $0.36$, $0.34$, $0.34$ and $0.30$ respectively). The excellent agreements between the model predictions and numerical simulations are shown in Fig.~\ref{fig:harddisksflat}, where one sees that small scale intermittency is successfully accounted for by the hard disk version of the vortex gas model.

\begin{figure}[ht]
%\hspace{0.0cm}
\includegraphics[width=0.49\textwidth]{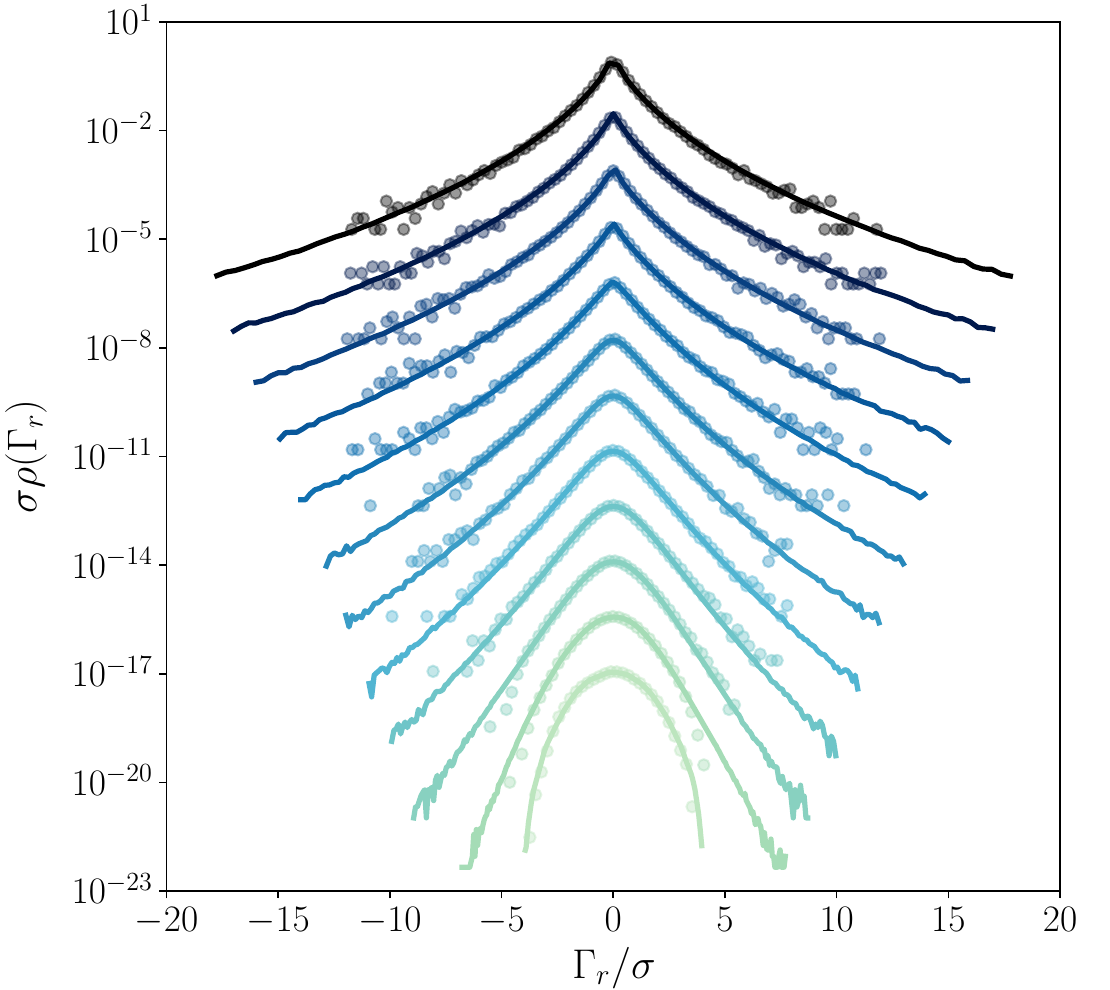}
%\vspace{0.0cm}
\caption{Standardized PDFs of circulation over square contours of sides $r=2^j\times(1.1\eta_K)$, with $j$ ranging from 0 to 10 (darker to lighter colors), for $R_\lambda=610$. Symbols: DNS. Lines: vortex gas model with hard disks. Curves are vertically shifted for clarity.}
\label{fig:pdfs}
\end{figure}

In Fig.~\ref{fig:pdfs} we plot the cPDFs for contours with sides varying from $r = 1.1$ to $r = 1126.4$ (in $\eta_K$ units), in the case $R_\lambda=610$. The sharp agreement between the DNS and the modeled cPDFs shows that a full multiscale description of circulation statistics has been achieved as well.

These compelling results indicate so far unnoticed aspects of the short-distance interactions between vortex structures. We, thus, proceed to investigate signatures of their apparent hard disk-like behavior from direct analyses of turbulent flow configurations, taken from the Johns Hopkins University DNS database \cite{JHTD,JHTD2}. The subject of vortex identification is a classic topic in the turbulence literature, specially active in wall bounded flows \cite{jeonghussain,zhouetal99,WuChristensen,chenetal18,chakrabortyetal05,chongetal90,huntetal88,elsasmoriconi,zhangetal18}. Here, we adopt the widely used swirling strength criterion \cite{zhouetal99,chakrabortyetal05} to identify vortex structures in
two-dimensional domains. Details about its implementation and validation procedures can be found in the Sec.~II of the Supplemental Material \cite{SM}.

We define, in the planar slices of a three-dimensional turbulent flow, a vortex center to be the point of maximum absolute vorticity inside the compact domain of each spotted vortex region. The vortex radius is furthermore estimated as the radius of a circle with an equivalent area. With this procedure, the detected vortex structures form a point process on the plane \cite{ripleybook} whose spatial statistics can be studied.
\begin{figure}[t]
%\hspace{0.0cm}
\includegraphics[width=0.49\textwidth]{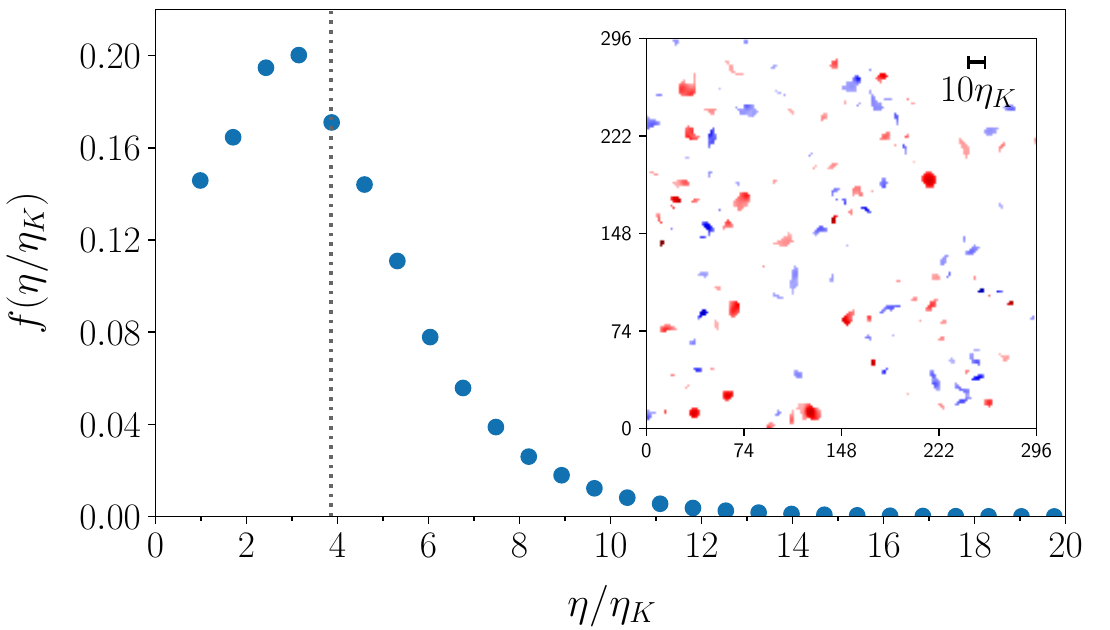}
%\vspace{0.0cm}
\caption{Distribution of estimated radii in DNS detected structures ($R_\lambda=610$). Vertical dotted line: the radius mean value $\langle\eta\rangle=3.85\eta_K$. Inset: a snapshot of the planar vortex spots (value axes labels given in units of $\eta_K$). Red and blue spots denote positive and negative vorticity, respectively.
}
\label{fig:radiuspdf}
\end{figure}
\begin{figure}[b]
%\hspace{0.0cm}
\includegraphics[width=0.49\textwidth]{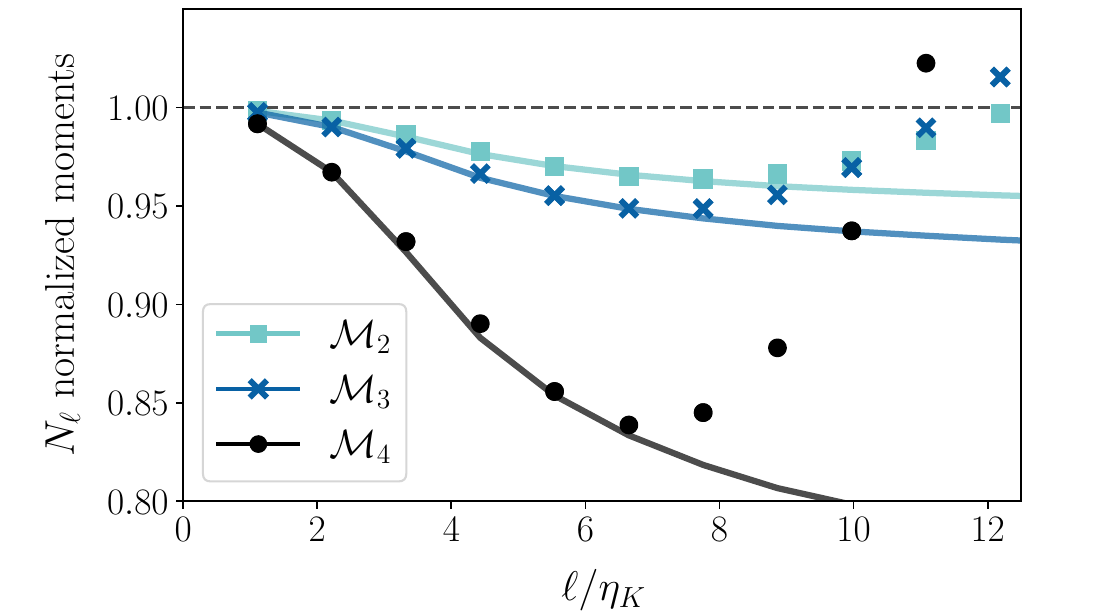}
%\vspace{0.0cm}
\caption{Normalized statistical moments of the number of points $N_\ell$ inside squares of side $\ell$ as a function of $\ell/\eta_K$. Symbols: DNS detected structures ($R_\lambda=610$). Solid lines: hard disk point process with the same mean number of points and radii $\langle \eta \rangle = 3.8 \eta_K$. Dashed: Poisson point process.}
\label{fig:spatialdist}
\end{figure}

In Fig.~\ref{fig:radiuspdf}, we show the estimated radius distribution obtained from DNS data with $R_\lambda=610$. The mean observed value is $\langle\eta\rangle=3.85 \eta_K$, while the distribution is peaked at $\eta_p = 3.15 \eta_K$,
which is in close agreement with the optimal radius found for the vortex gas model with hard disks, encoded in the parameter $a=3.3$.

To characterize the spatial distribution of structures, we rely on the statistical properties of $N_\ell$, the number of points (vortex centers) inside squares of side $\ell$. Some of its statistical moments are shown in Fig.~\ref{fig:spatialdist} for contours of variable sizes. Moments are normalized in order to have a constant unit value for a Poisson point process, and correspond to $\mathcal{M}_2 = \text{variance}[N_\ell]/\langle N_\ell\rangle$, $\mathcal{M}_3 = \text{skewness}[N_\ell]/\langle N_\ell \rangle^{-1/2}$ and $\mathcal{M}_4 = (\text{flatness}[N_\ell]-3)/\langle N_\ell \rangle^{-1}$.
Alongside DNS results (symbols), we show curves obtained with a gas of hard disks (known as a hard core Gibbs point process \cite{ripleybook}) containing the same mean number of points as the DNS and with a radius equivalent to $3.8\eta_K$ (that is, with a radius-to-box ratio of $3.8/L$ where $L$ is the DNS box size in $\eta_K$ units). Note that we have taken for comparison purposes the mean vortex radius $\langle \eta \rangle \simeq 3.8 \eta_K$ and not the peak value $\eta_p$
of the radius distribution. This follows from the fact that statistical signatures of hard-disk behavior are here related to the mutual exclusion between disks that may have completely different sizes.

The accurate modeling of DNS data reproduced in Fig.~\ref{fig:spatialdist} for small $\ell/\eta_K$ provides independent and clear evidence that the cores of close vortex structures tend not to overlap as they would do if they were Poisson distributed. As a consequence, intermittency growth is suppressed at small scales. It is important to emphasize that these are results of statistical nature, and are not inconsistent with the singular dynamical situations where vortex cores interact strongly, as in reconnection events \cite{hussain_durai}. We also add that the deviations observed in Fig.~\ref{fig:spatialdist}, between hard disk modeling and vortex configurational correlations for larger $\ell/\eta_K$, should be interpreted with care. They only mean that vortex structures develop further correlations at larger scales, reflected in the vortex gas model through the specific definitions provided by Eq.~(\ref{gamma_3}).

To summarize, we have improved and validated a vortex gas model of turbulent circulation, relying on explorations of DNS data and comparative Monte Carlo analyses. The model yields, in a very economical way (just a few parameters) and for a wide range of Reynolds numbers, a comprehensive description of the statistical properties of circulation fluctuations across the turbulent cascade scales. It turns out, as a statistical phenomenon, that turbulent vortex structures avoid each other at short distances, and this is found to be fundamentally connected to a proper account of small scale circulation intermittency.

From a methodological point of view, our approach renders clear that vortex identification methods and two-dimensional ``tomographic" cuts of three-dimensional structures (analogous to the ones commonly produced through the application of optical techniques in experimental fluid dynamics \cite{adrian_etal}) are instrumental tools of great heuristic relevance for the analysis of DNS data and the formulation of turbulence models. We have markedly benefited from them to establish links between turbulence, the theory of random point processes and classic models of statistical mechanics.

Interesting work is ahead, regarding extensions of the vortex gas model. A possible bridge to the scaling properties of velocity structure functions is worth investigating. We also draw particular attention to the problem of circulation statistics in non-planar contours and its connections to minimal surface theory \cite{Iyer_etal2,migdal}.

The vortex gas model presented here paves the way for deeper investigations on some of the most fundamental issues of turbulence, to the extent that its multifractal and structural elements are put together into a consistent and simple unifying picture. These have been invariably introduced as the bases of two disconnected descriptions of turbulent intermittency, a point of puzzling phenomenological understanding over the years.
\begin{acknowledgments}
We thank Gabriel Apolin\'ario for enlightening discussions. This study was financed in part by the Coordenação de Aperfeiçoamento de Pessoal de Nível Superior - Brasil (CAPES) - Finance Code 88887.336246/2019-00.
\end{acknowledgments}

%\end{document}
\clearpage

\newpage
\onecolumngrid
\appendix

\setcounter{equation}{0}
\setcounter{figure}{0}
\setcounter{table}{0}
\setcounter{page}{1}
\makeatletter
\renewcommand{\theequation}{S\arabic{equation}}
\renewcommand{\thefigure}{S\arabic{figure}}
\renewcommand{\thetable}{S\arabic{table}}
\setlength{\tabcolsep}{15pt}

\section*{{\large{Supplemental Material:}}\\
{\large{Circulation Statistics and the Mutually Excluding \\ Behavior of Turbulent Vortex Structures}}}
%\begin{center}
%L. Moriconi$^{1}$, R.M. Pereira$^{2}$, and V.J. Valadão$^{1}$\\
%\mbox{~}\\
%\textit{$^{1}$Instituto de F\'\i sica, Universidade Federal do Rio de Janeiro, \\
%C.P. 68528, CEP: 21945-970, Rio de Janeiro, RJ, Brazil}\\		
%\textit{$^{2}$Instituto de Física, Universidade Federal Fluminense, 24210-346 Niterói, RJ, Brazil}\\
%\end{center}
\vspace{0.2cm}

\subsection*{I. Subdominant Corrections for the Vortex Gas Model}

For the sake of analytical simplicity, take the domain $\mathcal{D}$ in Eq.~(5) of the main text to be a circle of radius $R$. 
In order to explore the asymptotic region $R/\eta_K\ll1$, we derive, from Eq.~(8) in the main text, the short distance expansion, 
\be\label{corr}
\langle\tilde{\omega}(\textbf{x})\tilde{\omega}(\textbf{y}) \rangle\approx 1 -\frac{|\textbf{x}-\textbf{y}|^2}{2\Delta^2}+\mathcal{O}(|\textbf{x}-\textbf{y}|^4)\approx e^{-\frac{|\textbf{x}-\textbf{y}|^2}{2\Delta^2}} \equiv g_\Delta(|\textbf{x}-\textbf{y}|),
\ee
where $\Delta^2=c^2\eta_K^2$ with $c^2=9/14$. Substituting in Eq.~(5) the continuous representation of the vortex surface density as $\sigma (\bfr) = \bar \sigma + \phi(\bfr) $, and working in the dilute gas approximation for a Poissonian distribution of vortices, the correlation functions of the fluctuating density field $\phi(\bfr)$ are, up to fourth order, given by \cite{PRE20sm}
\bea
&&\langle \phi(\bfr_1 ) \phi(\bfr_2) \rangle = \bar \sigma \delta_{12}, \\
&&\langle \phi(\bfr_1 ) \phi(\bfr_2) \phi(\bfr_3) \rangle = \bar \sigma \delta_{12} \delta_{13}, \\
&&\langle \phi(\bfr_1 ) \phi(\bfr_2) \phi(\bfr_3) \phi(\bfr_4) \rangle = \bar \sigma \delta_{12}\delta_{13}\delta_{14} + \bar \sigma^2 ( \delta_{12} \delta_ {34} + \delta_{13} \delta_ {24} + \delta_{14} \delta_ {23} ),
\eea
where $\delta_{ij}\equiv\delta^3(\textbf{x}_i-\textbf{x}_j)$. Up to second order in $\bar\sigma$, we get, from the above definitions, the circulation's variance
\be\label{varapp}
\langle \Gamma^2(\mathcal{D})\rangle= \langle\xi_\mathcal{D}^2\rangle
 \lr{\pi R^2}^2
 \lr{f_0-f_1\frac{R^2}{\eta^2_K}}  N_\eta
+\mathcal{O}\lr{\frac{R}{\eta_K}}^{8},
\ee
with
\be
f_0=
1+4N_\eta\left(
\frac{c^2}{c^2+2a^2}
\right) \label{f0}
\ee
and
\be
f_1=\frac{1}{4 a^2}+2N_\eta
\left(\frac{c^2}{(c^2+2a^2)^2}\right) \label{f1}
,
\ee
where $N_\eta=\bar{\sigma} \pi\eta^2_K a^2$ is the mean number of vortices in a disk of radius $\eta=a\eta_K$. Recall that the lognormal properties of $\xi_\mathcal{D}$ are described through OK62 phenomenology, Eqs.~(3) and (4) in the main text. Eqs.~(\ref{varapp}-\ref{f1}) have the flavor of a virial expansion, which is supposed to be meaningful if the vortex gas is sufficiently dilute. 
In addition, after a long and direct string of calculations based on the Gaussianity of $\tilde \omega(\bfr)$ (which allows us to factorize its higher order correlation functions into products of two-point correlators), 
we obtain the fourth order circulation's statistical moment,
\be
\langle \Gamma^4(\mathcal{D})\rangle=
\langle \xi^4_\mathcal{D}\rangle N_\eta \lr{\pi R^2}^4
\left(
f_2-f_3\frac{R^2}{\eta_K^2}
\right)+\mathcal{O}\lr{\frac{R}{\eta_K}}^{12}, \label{moment4gamma}
\ee 
where,
\be
f_2=\frac{3}{2}+N_\eta\left(
3+\frac{48c^2}{4a^2+3c^2}+\frac{6c^2}{2a^2+c^2}
\right),
\ee
and
\be
f_3=\frac{9}{8 a^2}+\frac{N_\eta}{a^2}\left(
\frac{3}{2}+72c^2
\frac{c^2+2a^2}{(4a^2+3c^2)^2}+
6c^2
\frac{c^2+3a^2}{(2a^2+c^2)^2}
\right) \ . \ 
\ee
A key point in the vortex gas model is that the circulation's flatness can be used to find how the mean surface vortex density $\bar \sigma$ depends upon the Reynolds number. In fact, from 
(\ref{varapp}) and (\ref{moment4gamma}), we get
\be\label{kurtapp}
\frac{\langle \Gamma^4(\mathcal{D})\rangle}{\langle \Gamma^2(\mathcal{D})\rangle^2}\approx
\frac{\langle \xi_\mathcal{D}^4\rangle}{\langle \xi_\mathcal{D}^2\rangle^2}
\frac{f_0^{-2}f_2}{N_\eta}\left[
1-\left(\frac{f_3}{f_2}-2\frac{f_1}{f_0}\right)\left(\frac{R}{\eta_K}\right)^2
\right]+\mathcal{O}\left(\frac{R}{\eta_K}\right)^{4} \ , \
\ee
which, when applied to Eq.~(6) in the main text, leads to
\be
\frac{f_2}{f_0^{2}N_\eta}=15^{3\mu/4} C_4 R_{\lambda}^{c_4-3\mu/2}\equiv F_0 \ , \
\ee
or, equivalently, to the cubic equation for $N_\eta$,
\be\label{poly}
F_0c_1^2N_\eta^3+2F_0c_1N_\eta^2+(F_0-c_2)N_\eta-\frac{3}{2}=0 \ , \
\ee
where
\be\label{c1}
c_1=\frac{4c^2}{c^2+2a^2} \ , \
\ee
and
\be\label{c2}
c_2=3+\frac{48c^2}{4a^2+3c^2}+\frac{6c^2}{2a^2+c^2} \ . \
\ee
The Poissonian modeling result, Eq.~(7) in the main text, is recovered here by setting $c_1=c_2=0$ in Eq.~(\ref{poly}).

As it is clear from Eq.~(\ref{moment4gamma}), the quantity $f_3/f_2-2f_1/f_0$, which depends on the parameter $a$, controls the curvature of the parabolic approximation for the flatness. Solving Eq.~(\ref{poly}) for the reference value $a=3.3$ \cite{PRE20sm}, we get $\bar{\sigma}$ which is about $5.5$ times larger than the original Poissonian (first order) evaluation. This is not a physically acceptable estimate, once it would contradict, against observations \cite{orszag_etalsm,farge_etalsm}, the diluted vortex gas approximation. Actually, as an illustrative account on the effects of subleading corrections, we note that the mean planar intervortex distance would be given by $\bar{\sigma}^{-1/2}\approx1.57\eta$ for $R_\lambda=240$, so that vortex structures would considerably overlap in this case.
As discussed in the main text, a phenomenological solution of this puzzling state of affairs is accomplished on the grounds that
the planar vortex structures are not Poisson distributed at dissipative scales, but rather distributed as a gas of hard disks.

% \section{Temporária - diferenças relativas:}

% Em unidades apropriadas:
% \be
% \langle\Gamma^2\rangle=(\pi R^2)^2\left(1-K_2 R^2\right)+\mathcal{O}\left(R^8\right)
% \ee
% \be
% \langle\Gamma^4\rangle=\frac{(\pi R^2)^4}{N_\eta f_0^2}\left(1-K_4 R^2\right)+\mathcal{O}\left(R^{12}\right)
% \ee
% \be
% F=\frac{\langle\Gamma^4\rangle}{\langle\Gamma^2\rangle^2}=C_4R_\lambda^{c_4}\left(1-(K_4-2K_2) R^2\right)+\mathcal{O}\left(R^4\right)
% \ee
% \be
% K_2(a,N_\eta)=\frac{f_1}{f_0}
% \ee
% \be
% K_4(a,N_\eta)=\frac{f_3}{f_2}
% \ee
% $f_i$s em geral possuem todas as ordens em $N_\eta$, ou seja, $\bar \sigma$ mas no point vortex nós consideramos $f_i(N_\eta)=c_i+\mathcal{O}(\bar{\sigma})$ e os calculos que fiz consideram $f_i(N_\eta)=c_i+d_i N_\eta+\mathcal{O}(\bar{\sigma}^2)$, definindo-se a diferença relativa:
% \be
% \delta F =\frac{F(d_i=0)-F(d_i)}{F(d_i=0)}
% \ee
% temos, para $a=3.3$ e $N_\eta=0.39$, que são os parametros do point vortex, temos diferenças nas funções $f_i$'s
% \be
% \delta f_0\approx -4.5\%
% \ee
% \be
% \delta f_1\approx -4.3\%
% \ee
% \be
% \delta f_2\approx -100\%
% \ee
% \be
% \delta f_3\approx -78\%
% \ee
% e diferenças nos observáveis,
% \be
% \delta K_2\approx 0.12\%
% \ee
% \be
% \delta K_4\approx 11\%
% \ee
% \be
% \delta K_f=\delta(K_4-2K_2)\approx 32\%
% \ee
% \be
% \delta \xi_0\approx 0.12\%
% \ee
% \be
% \delta \bar{\sigma}=\delta N_\eta\approx -226\%
% \ee
% OBS: essas informações podem vir em tabelas se necessário para a discussão

\subsection*{II. Vortex Structures and Circulation Statistics from DNS Data}

To detect planar intersections of turbulent vortex structures, we apply the swirling strength criterion \cite{zhouetal99sm,chakrabortyetal05sm} to two-dimensional slices of the velocity field obtained from direct numerical simulations (DNS) of the Navier-Stokes equations. The criterion consists in computing the eigenvalues of the velocity gradient tensor, $(\bfnabla\bfv)_{ij} = \partial v_i/\partial x_j$, and associating vortices to regions where the imaginary part of an eigenvalue is non-zero. Of course, since $\bfnabla\bfv$ is a $3\times 3$ tensor, if one eigenvalue is complex then it's complex conjugate also is, and the absolute value of their imaginary part, $\lambda_{ci}$, is called the local swirling strength. The rationale behind the criterion is that the evolution of tracers following frozen streamlines, when linearized around the origin, reads $\dot x_i = (\bfnabla\bfv)_{ij} x_j$, which shows spiraling orbits when $\bfnabla\bfv$ has complex eigenvalues. The criterion then defines vortex structures as domains where the $\lambda_{ci}$ field is non-zero. In fact, a threshold on $\lambda_{ci}$ is often employed to avoid noise and get smoother results. Following \cite{WuChristensensm,chenetal18sm}, we adopt the threshold $\lambda_T=1.5\sigma_{\lambda_{ci}}$, where $\sigma_{\lambda_{ci}}$ is the standard deviation of $\lambda_{ci}$. In this way, a connected region where $\lambda_{ci}>\lambda_T$ is considered a vortex. The method is applied to slices of a $R_\lambda=610$ simulation, resolved on 4096$^3$ grid points, available at the Johns Hopkins University DNS databases \cite{JHTDsm,JHTD2sm}. We use the single snapshot provided, taking samples of 64 equally spaced planar slices normal to each of the Cartesian axes, for a total of 192 planes.    

\begin{figure}[ht]
\includegraphics[width=0.97\textwidth]{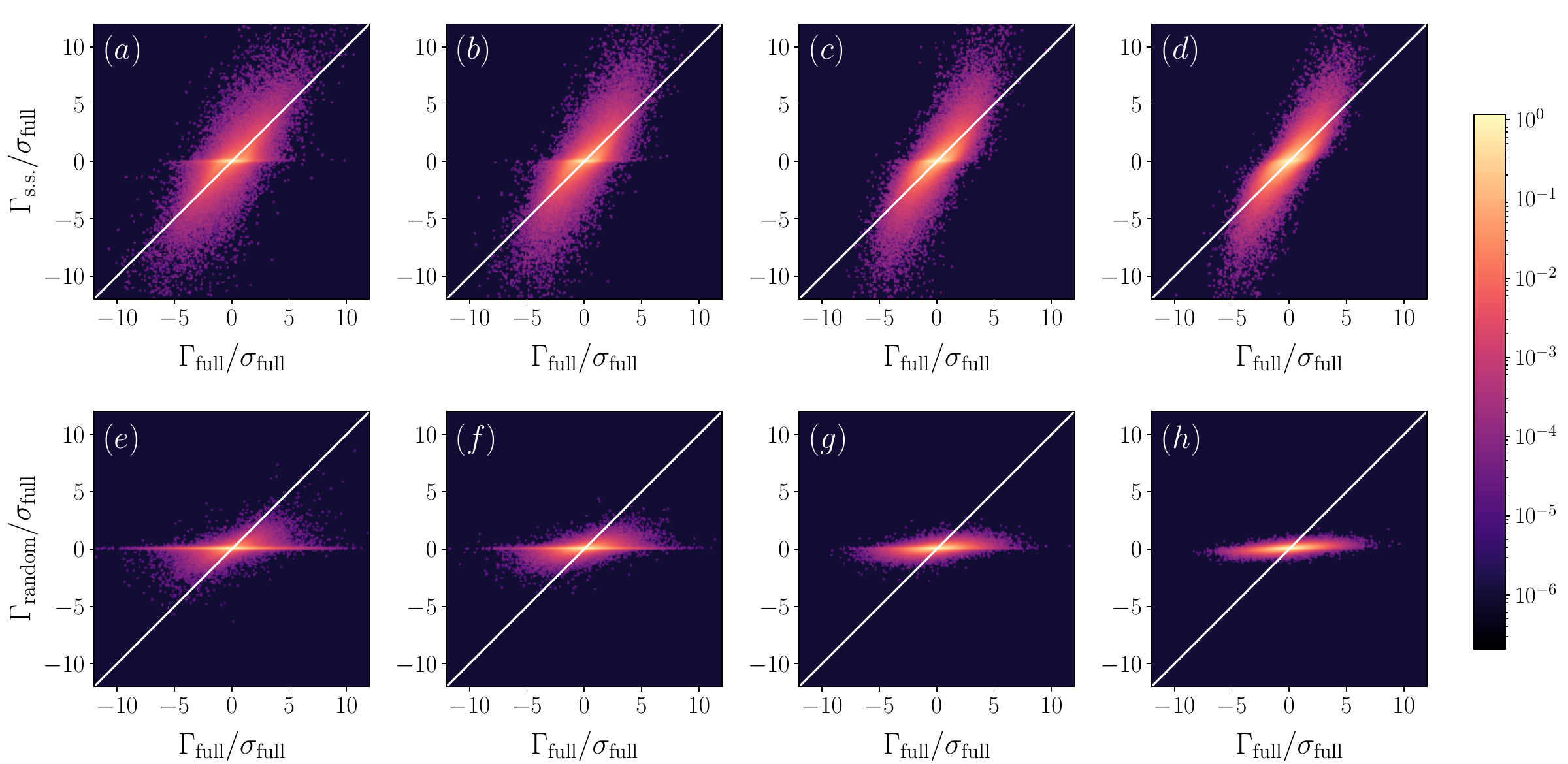}
\caption{Joint density plots for the circulation pairs
($\Gamma_\mathrm{full}$, $\Gamma_\mathrm{s.s.}$), panels (a-d), and
($\Gamma_\mathrm{full}$, $\Gamma_\mathrm{random}$), panels (e-h), over square contours positioned on the grid. All the circulations values are normalized by the full circulation's standard deviation $\sigma_\mathrm{full}$. Contours sides correspond, from left to right, to $10\eta_K$, $19\eta_K$, $37\eta_K$ and $72\eta_K$}
\label{fig:spots}
\end{figure}

A fundamental hypothesis of the vortex gas model is that vortex structures are responsible for velocity circulation fluctuations. We test this idea by combining DNS data and the model's main definitions as follows. A two-dimensional vorticity field is created, representing a DNS slice, by adding Gaussian functions $g_\eta(\bfr - \bfr_i)$ centered on every position $\bfr_i$ where a vortex was detected on the slice. Each Gaussian has the estimated vortex radius $\eta$ as its width and the maximum vorticity inside the vortex as its amplitude. We recall that, as explained in the main text, the radius of a detected structure is defined as the radius of a circle covering the same area. We then compare the circulation around square contours when computed from this swirling strength originated vorticity field ($\Gamma_\mathrm{s.s.}$) with those computed from the full DNS field ($\Gamma_\mathrm{full}$). In Figs.~\ref{fig:spots}(a-d) we show the obtained joint density plots for contours with increasing sides. %(a) $10\eta_K$, (b) $19\eta_K$ (c) $37\eta_K$ and (d) $72\eta_K$.
Abcissae correspond to $\Gamma_\mathrm{s.s.}$, while ordinates to $\Gamma_\mathrm{full}$. Data from all 192 DNS slices are plotted together and the white lines $y=x$ serve as reference. One sees a grouping tendency around the lines, showing that circulation is well captured by the combined and independent contributions of identified vortex structures.

To verify that this strong correlation is not a product of a simple generic sampling of the vorticity field, and hence that the detected structures are indeed the main actors in play, we now build a different vorticity field to serve as a null hypothesis test. First we pick random positions (Poisson distributed) on the DNS slices, saving the normal component of the vorticity at each one. This is done for 27564 points per slice, which is the mean number of detected structures on a DNS slice. Then, Gaussian functions centered on those positions are superposed, using the measured vorticities as amplitudes and the mean detected vortex radius, $\langle \eta\rangle = 3.85\eta_K$, as width. Joint density plots of Figs.~\ref{fig:spots}(e-h) once again compare $\Gamma_\mathrm{full}$ with the circulation computed with this newly constructed field $\Gamma_\mathrm{random}$, for the same contours as before. The lack of correlation concludes that structures created randomly from the vorticity field do not account for the circulation fluctuations. In particular, it is seen that large fluctuations are completely suppressed.

To further quantify these correlations, we compute for each one of the cases reported in Fig.~\ref{fig:spots} the Pearson coefficient,
\be
\rho_{XY} = \frac{\langle \left(X-\langle X \rangle \right) \left(Y- \langle Y \rangle \right) \rangle}{\sigma_X\sigma_Y} \ , \
\ee
where $\sigma_X$ and $\sigma_Y$ are the standard deviations of the random samples $X$ and $Y$, respectively.
Results, summarized in table \ref{tab:pearson}, further point to strong correlations between the circulation observed in DNS with that measured from the field associated to structures detected with the swirling strength criterion. Correlations are, in contrast, significantly reduced when structures randomly sample the flow.
\begin{table}[h]
\begin{tabular}{r||c|c|c|c}
contour sides  & $10\eta_K$ & $19\eta_K$ & $37\eta_K$ &  $72\eta_K$ \\ \hline
$(\Gamma_\mathrm{full},\Gamma_\mathrm{s.s.})$ & 0.70 & 0.73 & 0.76 & 0.81\\ 
$(\Gamma_\mathrm{full},\Gamma_\mathrm{random})$ & 0.32 & 0.39 & 0.45 & 0.52
\end{tabular}
\caption{Pearson correlation coefficients between $\Gamma_\mathrm{full}$ and both $\Gamma_\mathrm{s.s.}$ and $\Gamma_\mathrm{random}$ over contours of varying sides.}
\label{tab:pearson}
\end{table}

\end{document}